\documentclass[showpacs,fleqn,nobibnotes]{revtex4}

\usepackage{amsmath}
\usepackage{graphicx}
\usepackage{float}
\usepackage{subfigure}

\def\lsim{\raise0.3ex\hbox{$<$\kern-0.75em\raise-1.1ex\hbox{$\sim$}}}
\def\gsim{\raise0.3ex\hbox{$>$\kern-0.75em\raise-1.1ex\hbox{$\sim$}}}

\begin{document}

\title{Nuclear shadowing from exclusive quarkonium photoproduction at the BNL RHIC and CERN LHC}
\author{A. L. Ayala Filho, V.P. Gon\c{c}alves and M. T. Griep }

\affiliation{ Instituto de F\'{\i}sica e Matem\'atica, Universidade Federal de
Pelotas\\
Caixa Postal 354, CEP 96010-900, Pelotas, RS, Brazil.}

\begin{abstract}
The photonuclear production of vector mesons in ultraperipheral heavy ion collisions is investigated within the 	 collinear approach using different parameterizations for the nuclear gluon distribution. The integrated cross section and the rapidity distribution for the $AA \rightarrow V\,AA$ ($V = J/\Psi,\,\Upsilon$) process are computed for energies of RHIC and LHC. A comparison with the recent PHENIX data on coherent production of $J/\Psi$ mesons is also presented. We demonstrate that the study of the exclusive quarkonium photoproduction  can be used to constrain the nuclear effects in the gluon distribution.

\end{abstract}

\maketitle

A systematic measurement of the nuclear
gluon distribution is of fundamental interest in understanding the parton structure of nuclei and to determine the initial
conditions of the quark gluon plasma (QGP) predicted to be formed in central heavy ion collisions (See e.g. \cite{qgp}). Another important motivation for the
determination of the nuclear gluon distribution is that the high
density effects expected to occur in the high energy limit of QCD
should be manifest in the modification of the gluon dynamics \cite{jamal}.
At the moment the behavior
of this distribution is completely undetermined by the fixed
target experiments, with a possible improvement in the future
electron-nucleus colliders (See e.g. \cite{raju,vic_simone}). However, as the date of construction
and start of operation of these colliders is still in debate, we
need to obtain  alternative searches to estimate the medium
effects in the nuclear gluon distribution.
It has motivated several authors to propose the study of different observables in distinct processes to constrain the nuclear gluon distribution (For a recent review see Ref. \cite{armesto}). In particular,  in Ref. \cite{vic_ber} it was proposed to study the  vector meson production in ultraperipheral heavy ion collisions at RHIC and LHC in order to constrain the nuclear medium effects present in the nuclear gluon distribution. The basic idea is that in these collisions the high
flux of quasi-real photons from one of the nucleus provides a
copious source of photoproduced reactions \cite{bert,bert2,bert3,bert4}. A photon stemming
from the electromagnetic field of one of the two colliding nuclei
can penetrate into the other nucleus and interact with one or more
of its hadrons, giving rise to photon-nucleus collisions to an
energy region hitherto unexplored experimentally.  For example, the
interaction of quasi-real photons with protons has been studied
extensively at the electron-proton collider at HERA (For a recent review see Ref. \cite{ivanov_nik}). The obtained $\gamma p$ center of mass
energies extends up to $W_{\gamma p} \approx 300$ GeV, an order of
magnitude larger than those reached by fixed target experiments.
Due to the larger number of photons coming from one of the
colliding nuclei in heavy ion collisions, a  similar and more detailed
study will be possible in these collisions, with $W_{\gamma N}$
reaching 950 GeV for the Large Hadron Collider (LHC) operating in
its heavy ion mode \cite{bert,bert2,bert3,bert4}. As the cross section for the diffractive vector
meson production depends (quadratically) on the gluon
distribution, it gives a unique opportunity to study the low $x$
behavior of the gluons inside the nucleus. The  results from Ref. \cite{vic_ber} demonstrate
that the study of photoproduction of vector mesons in ultraperipheral collisions  determines the
behavior of the nuclear gluon distribution in the full kinematic
region, with the rapidity distribution allowing for the first time
to estimate the EMC, antishadowing and shadowing effects. However, these results were obtained considering several approximations for the calculations of the diffractive vector meson production, the GRV95 parameterization \cite{grv} for the nucleon parton distribution and the EKS parameterization \cite{EKS98} for the nuclear effects. All these models have been improved recently.
Furthermore,  the PHENIX Collaboration recently released the first (preliminary) data on the cross section of the coherent $J/\Psi$ production in gold - gold ultraperipheral collisions at $\sqrt{s} = 200$ GeV \cite{phenix}, providing the first opportunity to check the basic features and main approximations of the distinct approaches describing nuclear vector meson photoproduction. Finally, in near future the Large Hadron Collider (LHC) at CERN will
start its experimental physics program and coherent hadron-hadron interactions should be studied \cite{david}.
These facts motivate a revision  of the  estimates presented in Ref. \cite{vic_ber}. In particular, in this paper  we revise  the predictions for $J/\Psi$ production in coherent $AA$ collisions and present, for the first time, predictions for the $\Upsilon$ production in coherent collisions considering the collinear formalism.



Over the past few  years a comprehensive analysis of  vector meson  production in ultraperipheral heavy ion collisions was made considering different theoretical approaches \cite{vic_ber,klein_prc,strikman_plb,per4,klein_prl,strikman_jhep,vicmag_prd2008,ivanov_kop}. In particular,   much effort has been devoted to obtain signatures of the QCD Pomeron in such process \cite{per4,vicmag_prc,vicmag_prd2008}, which can be used to constrain the QCD dynamics at high energies.
In this paper  we restrict our study of vector meson production to the  collinear formalism, based on the QCD factorization theorem \cite{qcdfac}, which predicts
 that at small-$x$ and for a sufficiently hard scale, the cross section of coherent production of vector mesons off any hadronic target, including a nucleus, is proportional to the square of the gluon parton density of the target.
 To lowest order, the $\gamma h \rightarrow V  h$ $(h = p, \,A)$
amplitude can be factorized into the product of the $\gamma
\rightarrow Q \overline{Q}$ transition $(Q = c, \,b)$, the scattering of the
$Q\overline{Q}$ system on the target via (colorless) two-gluon
exchange, and finally the formation of the quarkonium from the
outgoing $Q\overline{Q}$ pair.  The heavy meson mass $M_{V}$
ensures that perturbative QCD can be applied to photoproduction.
The calculation was
performed some years ago to leading logarithmic ($\log(\overline{Q}^2)$) approximation,
assuming the produced vector-meson quarkonium system to be
nonrelativistic \cite{ryskin,brod} and improved in distinct aspects
\cite{levin,fran}.
 Assuming a non-relativistic wave-function for the vector meson one have that the $t=0$
differential cross section of photoproduction of heavy vector
mesons in leading order collinear approximation  is given by \cite{ryskin,brod}
\begin{eqnarray}
\frac{d\sigma(\gamma h \rightarrow V h)}{dt}|_{t=0} =
\frac{\pi^3 \Gamma_{ee} M_{V}^3}{48 \alpha}
\frac{\alpha_s^2(\overline{Q}^2)}{\overline{Q}^8} \times
[xg_h(x,\overline{Q}^2)]^2\,\,, \label{sigela}
\end{eqnarray}
where $xg_h$ is the target gluon distribution, $x =
4\overline{Q}^2/W^2$ with $W$ the center of mass energy and
$\overline{Q}^2 = M_{V}^2/4$. Moreover, $\Gamma_{ee}$ is the
leptonic decay width of the vector meson.
The strong dependence on $xg$ offers an opportunity to use the experimental HERA data for the $\gamma^*p \rightarrow J/\Psi p$ process to determine the behavior of the gluon distribution at low $x$ and in the $Q^2$ region,  which is not  constrained by the global analyzes \cite{teubner}. In Refs. \cite{levin,fran,teubner,martin_ups,ivanov_nlo} the authors have estimated the
relativistic corrections [${\cal{O}}(4\%)$] , the real part contribution of the production amplitude [${\cal{O}}(15\%)$], the effect of off-diagonal partons [${\cal{O}}(20\%)$] and next-to-leading order corrections [${\cal{O}}(40\%)$]
 to the LO exclusive heavy vector meson production, given by Eq. (\ref{sigela}). 
 

\begin{figure}[t]
\includegraphics[scale=0.37]{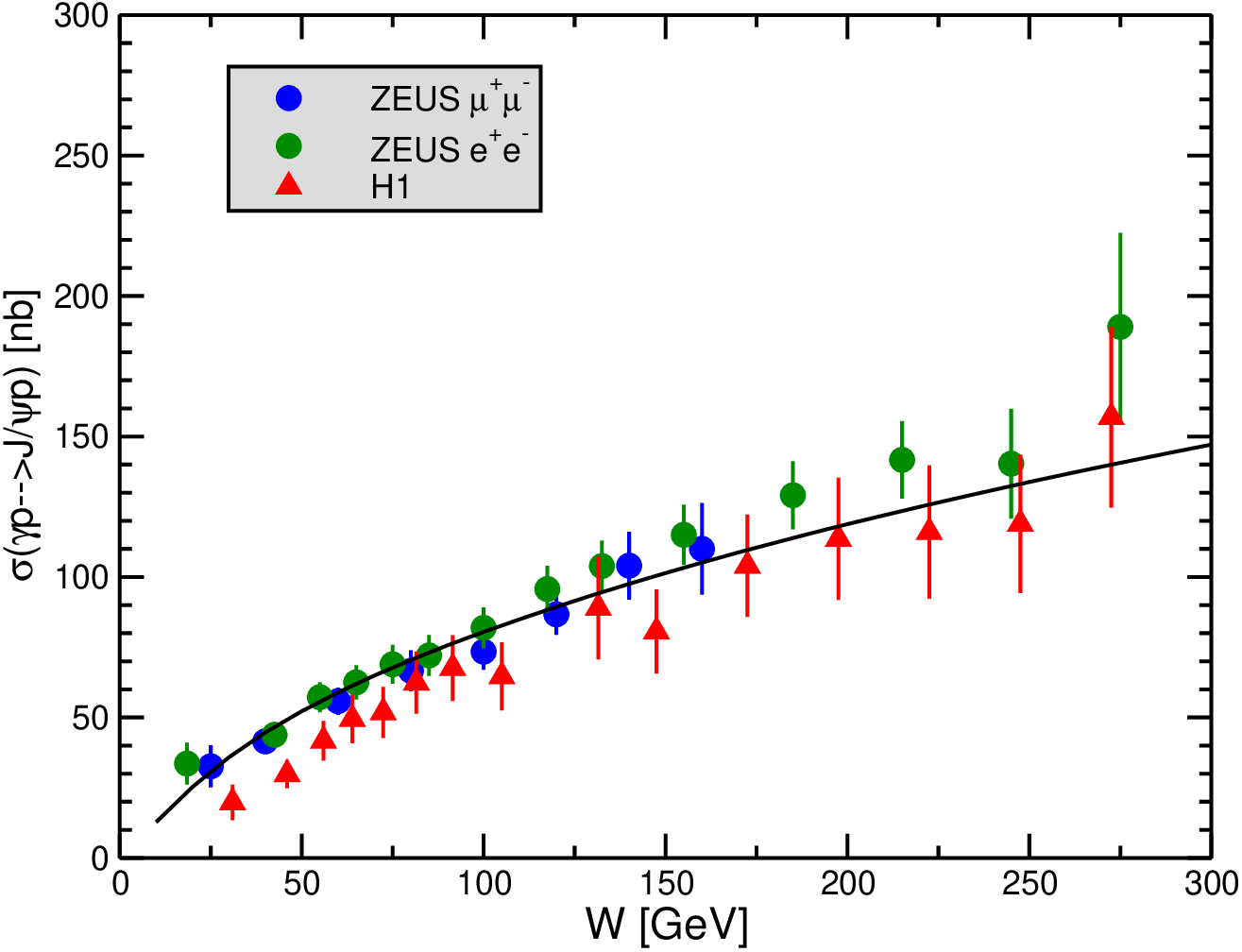}
\hspace{0.2cm}
\includegraphics[scale=0.37]{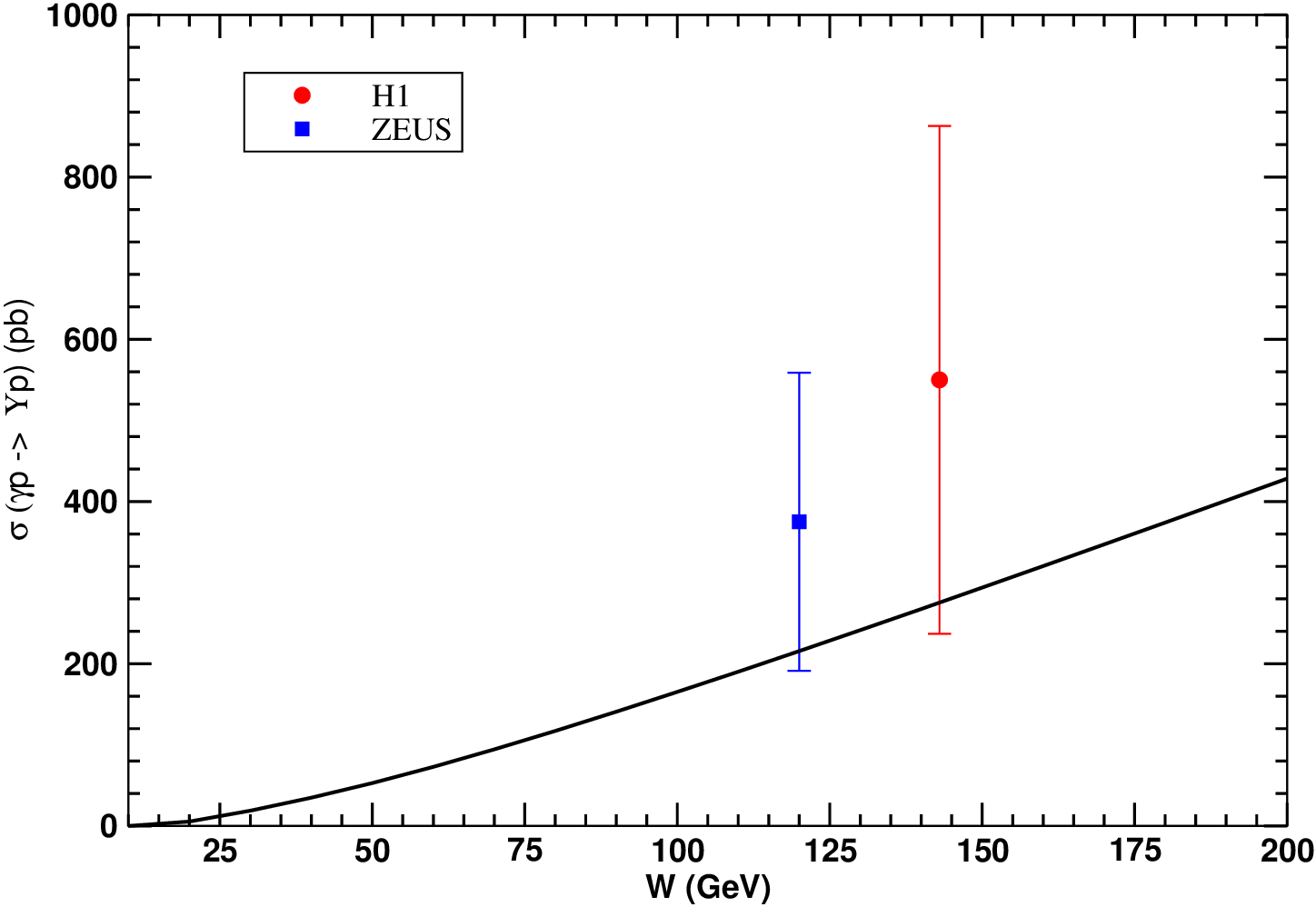}
\caption{(Color online) Energy dependence of the exclusive $J/\Psi$ (left panel) and $\Upsilon$ (right panel) photoproduction cross section. Comparison with HERA data \cite{H1_jpsi,ZEUS_jpsi,ZEUS_ups}.}
\label{fig1}
\end{figure}


In order to obtain a baseline for our calculations of the vector meson production in $\gamma A$ interactions we initially estimate the $\gamma p$ cross section and compare it with the HERA data.  Following Refs. \cite{martin_ups,teubner} we  estimate the total cross section for
 the $J/\Psi$ and $\Upsilon$ photoproduction at HERA using the MRST(LO) parameterization \cite{mrst} for the nucleon gluon distribution and including the corrections discussed above. Assuming
  an exponential parameterization for the small $|t|$
behavior of the amplitude, one have  that $
\sigma\, (\gamma p \rightarrow Vp) = \frac{1}{b} \frac{d\sigma(\gamma h \rightarrow V h)}{dt}|_{t=0}$, where $b$ is the experimentally measured slope parameter. In our calculations we assume $b = 4.5$ GeV$^{-2}$ as in Refs. \cite{levin,teubner}, which is in agreement with \cite{bh1}. For simplicity, we assume the same value of $b$ for $J/\Psi$ and $\Upsilon$ production. It is a reasonable approximation, since the HERA data shown that $b$ decreases to a universal value of  about 4 - 5 GeV$^{-2}$ as the scale $Q^2 + m_V^2$ increases, independently of the vector meson studied (For a recent review see, e.g., \cite{levy}).
   In Fig. \ref{fig1} we compare our results with the $J/\Psi$ (left panel) and $\Upsilon$ (right panel)  HERA data  \cite{H1_jpsi,ZEUS_jpsi,ZEUS_ups} and demonstrate that it are reasonably described. In the case of $\Upsilon$ production,  our  prediction is systematically somewhat below  the scarce experimental data. In principle, the agreement  could be improved by choosing a smaller value of $b$. 
In what follows we will assume that the corrections for the
LO exclusive heavy vector meson production are independent of the target, i.e. we consider that it are the same in the proton and nuclear case. Moreover, we assume that
in the case of nuclear targets, $b$ is dominated by the nuclear
size, with $b\sim R_A^2$  and the non-forward differential cross section is
dominated by the nuclear form factor, which is the Fourier
transform of the nuclear density profile. Therefore, we consider that the total cross section for vector meson production in $\gamma A$ interactions is given by
\begin{eqnarray}
\sigma(\gamma A \rightarrow V A) = \frac{d\sigma(\gamma A
\rightarrow V A)}{dt}|_{t=0} \, \int_{t_{min}}^{\infty} dt \,
|F(t)|^2 \,\,\,, \label{photonuc}
\end{eqnarray}
where $t_{min} = (M_{V}^2/2\omega)^2$ and $F(t)=\int d^3r \
\rho(r) \ \exp (i{\bf q}\cdot {\bf r})$ is the nuclear form factor
for the distribution. Here we use the
analytical approximation of the Woods-Saxon distribution as a hard
sphere, with radius $R_A$, convoluted with a Yukawa potential with
range $a=0.7$ fm. Thus, the nuclear form factor reads
as \cite{klein_prc},
\begin{equation}
F(q=\sqrt{|t|}) = \frac{4\pi\rho_0}{A\,q^3}\,
\left[\sin(qR_A)-qR_A\cos(qR_A)\right]
\,\left[\frac{1}{1+a^2q^2}\right]\,\,, \label{FFN}
\end{equation}
where $\rho_0 = 0.16$ fm$^{-3}$.

\begin{figure}[t]
\includegraphics[scale=0.5]{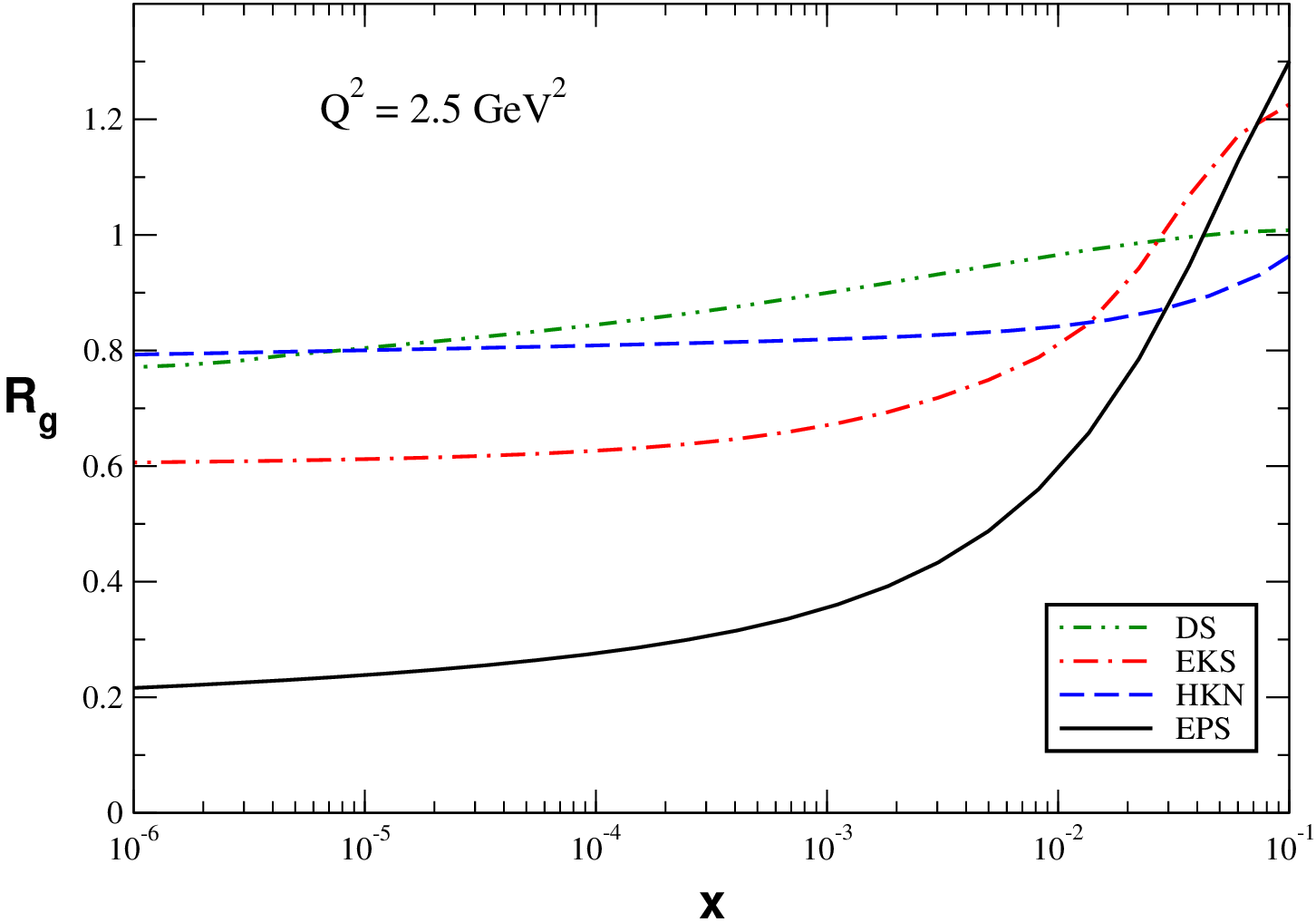}
\caption{(Color online)  Ratio $R_g \equiv xg_A/A.xg_N$ predicted by the  DS \cite{sassot}, EKS \cite{EKS98}, HKN \cite{HKN} and EPS \cite{EPS08} parameterizations at $Q^2 = 2.5$ GeV$^2$ and $A = 208$. }
\label{fig:rg}
\end{figure}

The main input in the  calculations of the quarkonium production cross section in photon-nucleus interactions is the nuclear gluon distribution. In the last years several  groups has proposed parameterizations for the nuclear parton distributions (nPDF), which are based on different assumptions and techniques to perform a global fit of different sets of
data using the DGLAP evolution equations \cite{EKS98,sassot,HKM,HKN,EPS08}. Lets present a brief review of these parameterizations (For details see, e.g., Ref. \cite{armesto}). Initially, let's consider the EKS parameterization \cite{EKS98} which  is based on the DGLAP evolution equations  \cite{dglap} with adjusted initial parton distributions to describe experimental results \cite{arneodo,e665} presenting nuclear shadowing effects. The basic idea of this framework is the same as in the global analyzes of parton distributions in the free proton: they determine the nuclear parton densities at a wide range of $x$ and $Q^2\geq Q_0^2=2.25$ GeV$^2$ through their perturbative DGLAP evolution
by using the available experimental data from $lA$ DIS and Drell-Yan (DY)
measurements in $pA$ collisions as constraint. In this approach, the nuclear
effects are taken into account in the initial parton distribution $xf^A(x,Q_0^2)$ of
the DGLAP evolution. EKS have expressed the
results in terms of the nuclear ratios $R_f^A(x,Q^2)$ for each parton flavor
$f$ in a nucleus with $A$ nucleons ($A>2$), at $10^{-6}\leq x\leq 1$ and $
2.25\,GeV^2\leq Q^2\leq 10^4\,GeV^2$.
Other groups have considered different sets of data,
next-to-leading order (NLO) corrections to the DGLAP equation
and uncertainty estimates, and have proposed distinct approaches to describe the nuclear effects \cite{sassot,HKM,HKN,EPS08}. Next-to-leading order (NLO) corrections to the DGLAP evolution are taken into account in \cite{sassot,HKN}. In particular, De Florian and Sassot (DS) \cite{sassot} proposed a framework where each nuclear parton distribution is described by a convolution of the corresponding free nucleon parton distribution with a simple flavor dependent weight function that takes into account the nuclear effects.
On the other hand, the authors of \cite{HKM,HKN} have calculated the nPDF uncertainties by using the Hessian method and have obtained LO \cite{HKM} and NLO \cite{HKN} nuclear parameterizations.
More recently, an important step toward better constraining the gluon nuclear distributions was done in \cite{EPS08}, where the authors include recent RHIC data from high-$p_T$ hadron production at forward rapidities in d+Au collisions, which probe small values of $x$ in the gluon distribution of the $Au$ target, in their global analysis. The striking result is a much stronger gluon shadowing than in the previous parameterizations.
Due to the scarce experimental data in the small-$x$ region and/or for observables strongly dependent on the nuclear gluon distribution, the current status is that its behavior is completely undefined. It is demonstrated by the analysis of the Fig. \ref{fig:rg}, where we present the results for the ratio  $R_g \equiv xg_A/A.xg_N$ predicted by the  EKS \cite{EKS98},  DS \cite{sassot}, HKN \cite{HKN} and EPS \cite{EPS08} parameterizations at $Q^2 = 2.5$ GeV$^2$ and $A = 208$.  As we can see, these parameterizations predict very distinct magnitudes for the nuclear effects. For larger values of $x$, the EKS and the EPS show antishadowing, while this effect is absent for the HKN and EPS parameterizations in the $x \le 10^{-1}$ domain. The more surprising feature is however the amount of shadowing in the different parameterizations. While the shadowing is moderate for DS and HKN parameterizations 
and somewhat
bigger for EKS one, the EPS prediction has a much stronger suppression compared with the other parameterizations. For smaller $x$ around $x\simeq {10}^{-5}$, while DS and HKN parameterizations have about $20\%$ suppression and EKS one have about $40\%$ suppression, for the EPS parameterization this effect goes to almost $80\%$ suppression in the nuclear gluon compared with the $A$ scaled gluon content in the proton! For bigger values of $x$ the behavior is distinct for all parameterizations. As $x$ grows, the DS parameterization predicts that $R_g$ grows continuously to 1, that means that the shadowing dies out when $x\to 10^{-1}$. The same happens for the HKN one in this limit, but this growth starts only at $x>10^{-2}$, with $R_g$ being flat for $10^{-5}<x< 10^{-2}$. At $x\approx 10^{-1}$, we have that behaviors predicted by the EKS and EPS parameterizations are similar, with $R_g$ exceeding 1.2. The main distinction between these parameterizations is that in the EPS parameterization  one has a much steeper growth, from a much stronger suppression at smaller $x$ to the antishadowing behavior for larger values of $x$.
The  difference between the distinct parameterizations observed in Fig. \ref{fig:rg} will be amplified in the exclusive quarkonium photoproduction   in $\gamma A$ interactions due to the quadratic dependence on $xg_A$ of the cross section. Consequently, it is expected a large difference between the predictions obtained using the distinct parameterizations. This is the main motivation for our calculations.


\begin{figure}[t]
\includegraphics[scale=0.5]{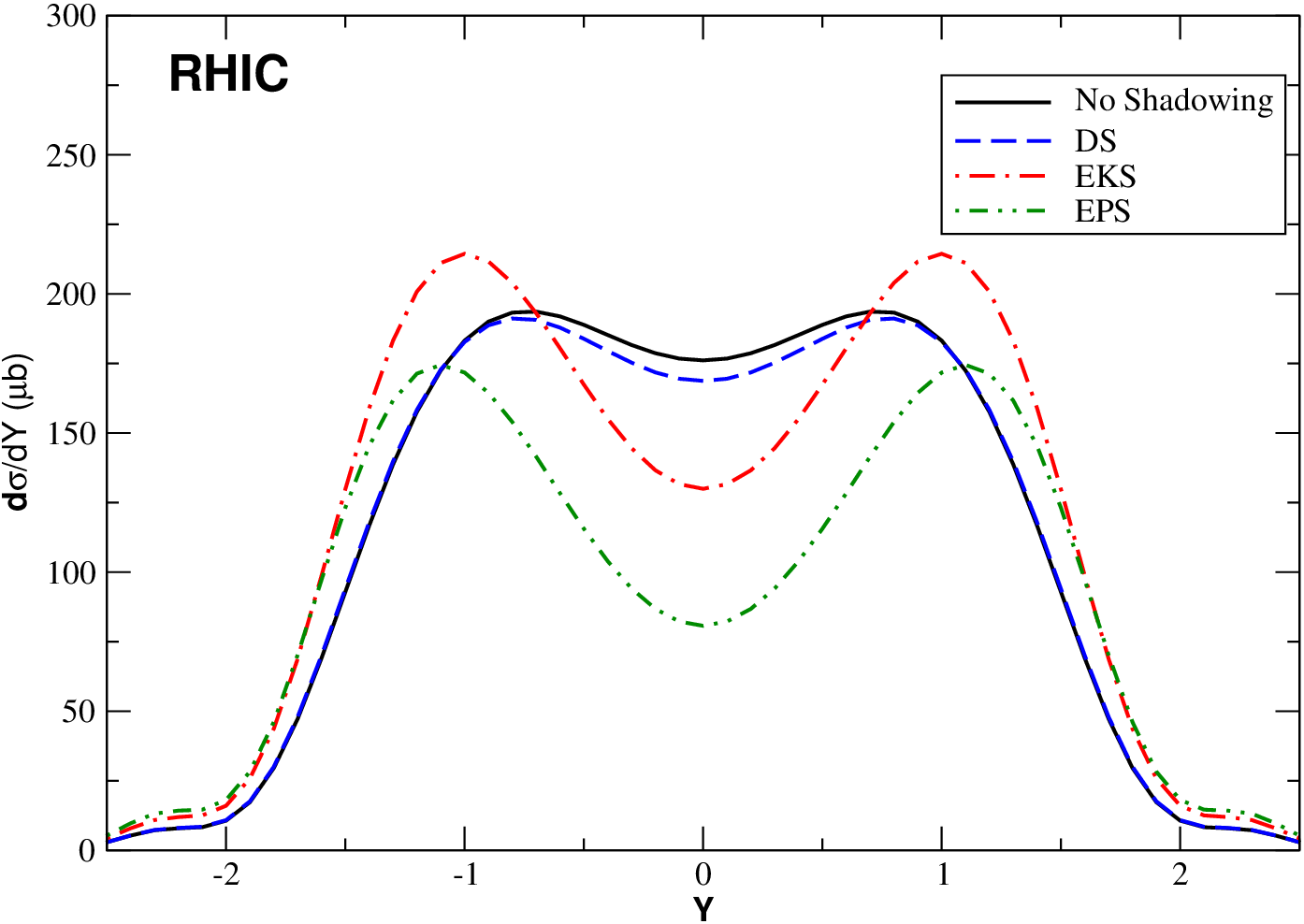}
\caption{(Color online) Rapidity distribution for $J/\Psi$ photoproduction in $AuAu$ collisions at RHIC.  }
\label{fig:rhic}
\end{figure}

We now present a brief review of vector meson production in ultraperipheral relativistic heavy-ion collisions (UPC's), introducing the main formulas. In UPC's the ions do
not interact directly with each other and move essentially
undisturbed along the beam direction. The only possible
interaction is due to the long range electromagnetic interaction
and diffractive processes (For a review see, e. g. Refs.
\cite{bert,bert2,bert3,bert4}). In the particular case of photonuclear reactions  the cross sections are given by the convolution between
the photon flux from one of the nuclei and the cross section for
the scattering photon-nuclei. The final  expression
for the production of vector mesons in ultraperipheral heavy ion
collisions is then given by,
\begin{eqnarray}
\sigma_{AA \rightarrow AAV}\,\left(\sqrt{S_{\mathrm{NN}}}\right) = \int \limits_{\omega_{min}}^{\infty} d\omega \, \frac{dN\,(\omega)}{d\omega}\,\, \sigma_{\gamma \,A \rightarrow VA} \left(W_{\gamma A}^2=2\,\omega\sqrt{S_{\mathrm{NN}}}\right)\,
\label{sigAA}
\end{eqnarray}
where $V = J/\Psi$ or $\Upsilon$, $\omega$ is the photon energy  with $\omega_{min}=m_V^2/4\gamma_L m_p$ and
$\sqrt{S_{\mathrm{NN}}}$ is  the ion-ion c.m.s energy. The Lorentz factor for LHC is
$\gamma_L=2930$, giving the maximum c.m.s. $\gamma N$ energy
$W_{\gamma A}  \lesssim 950$ GeV.  In this process  the nuclei are not disrupted and the final state consists solely of the two nuclei and the vector meson decay products. Consequently,  the final state is  characterized by a small number of centrally produced particles, with rapidity gaps separating the central final state from both beams. Moreover, due to the coherence requirement, the transverse momentum  is limited to be small. Therefore, these reactions can be studied experimentally by selecting events with low multiplicity and small total $p_T$. It is important to emphasize that recent STAR $\rho^0$ photoproduction data \cite{star_rho} show significant coherent production to $p_T \le 150$ MeV. 
The photon flux is given by the
Weizsacker-Williams method \cite{bert}. The flux from a charge
$Z$ nucleus a distance $b$ away is
\begin{eqnarray}
\frac{d^3N\,(\omega,\,b^2)}{d\omega\,d^2b}= \frac{Z^2\alpha_{em}\eta^2}{\pi^2 \,\omega\, b^2}\, \left[K_1^2\,(\eta) + \frac{1}{\gamma_L^2}\,K_0^2\,(\eta) \right] \,
\label{fluxunint}
\end{eqnarray}
where $\gamma_L$ is the Lorentz boost  of a single beam and $\eta
= \omega b/\gamma_L$; $K_{0,\,1}(\eta)$ are the
modified Bessel functions. The requirement that  photoproduction
is not accompanied by hadronic interaction (ultraperipheral
collision) can be done by restricting the impact parameter $b$  to
be larger than twice the nuclear radius, $R_A=1.2 \,A^{1/3}$ fm.
Therefore, the total photon flux interacting with the target
nucleus is given by Eq. (\ref{fluxunint}) integrated over the
transverse area of the target for all impact parameters subject to
the constraint that the two nuclei do not interact hadronically.
An analytic approximation for $AA$ collisions can be obtained
using as integration limit $b>2\,R_A$, producing
\begin{eqnarray}
\frac{dN\,(\omega)}{d\omega}= \frac{2\,Z^2\alpha_{em}}{\pi\,\omega}\, \left[\bar{\eta}\,K_0\,(\bar{\eta})\, K_1\,(\bar{\eta})+ \frac{\bar{\eta}^2}{2}\,\left(K_1^2\,(\bar{\eta})-  K_0^2\,(\bar{\eta}) \right) \right] \,
\label{fluxint}
\end{eqnarray}
where $\bar{\eta}=2\omega\,R_A/\gamma_L$.
It is worth mentioning that the
difference between the complete numeric and the analytical
calculation presented above  for the photon flux is less than 15
\% for the most of the purposes \cite{bert,bert2,bert3,bert4}.

In what follows we
calculate the rapidity distribution and total cross sections for quarkonium production  in ultraperipheral heavy ion collisions at RHIC and LHC energies. Moreover, since none of the RHIC and LHC detectors have full rapidity coverage, we also present our predictions for $d\sigma/dY$ at $Y = 0$. We assume that $xg_A (x,Q^2) = R_g (x,Q^2) . A \, xg_p(x,Q^2)$, with $R_g$ given by the DS, EKS and EPS parameterizations  and $xg_p$ given by the MRST(LO) parameterization \cite{mrst}. As the HKN parameterization is similar to the DS one in the kinematical range considered, it is not included in our analyses.
The distribution on rapidity $Y$ of the produced final state can be directly computed from Eq. (\ref{sigAA}), by using its  relation with the photon energy $\omega$, i.e. $Y\propto \ln \, (2 \omega/m_V)$.  Explicitly, the rapidity distribution is written down as,
\begin{eqnarray}
\frac{d\sigma \,\left[A + A \rightarrow   A \otimes V \otimes A \right]}{dY} = \omega \frac{dN_{\gamma} (\omega )}{d\omega }\,\sigma_{\gamma A \rightarrow V A}\left(\omega \right)\,
\label{dsigdy}
\end{eqnarray}
where $\otimes$ represents the presence of a rapidity gap.
Consequently, given the photon flux, the rapidity distribution is thus a direct measure of the photoproduction cross section for a given energy.
Initially we   present  our predictions for $J/\Psi$  production in coherent $AA$ collisions at RHIC, considering $A = Au$ and $\sqrt{s_{NN}} = 200$ GeV. The production at mid-rapidity at the RHIC probes $x$-values of order  $10^{-2}$, where there is a large difference between the  distinct nuclear parameterizations (See Fig. \ref{fig:rg}).
For comparison we also present the prediction obtained assuming $R_g = 1.0$ (no shadowing). Our results are shown in Fig. \ref{fig:rhic}. We have that the DS prediction is similar to the no shadowing one. This behavior is expected due to the small magnitude of the nuclear effects present in this parameterization at $x \gtrsim 10^{-2}$. On the other hand, the presence of the  antishadowing effect in the EKS and EPS parameterizations modify the rapidity distribution at large $|Y|$, implying an enhancement in the distribution.   The study of this kinematical range can be useful to constrain the magnitude of the antishadowing effects. In contrast, the behavior of the distribution at mid-rapidity is directly associated to the magnitude of the shadowing effect in the nuclear gluon distribution, which is larger in the EPS parameterization in comparison to the EKS one. It implies the larger suppression of the rapidity distribution at $Y = 0$ when calculated with the EPS gluon observed in Fig. \ref{fig:rhic}. These behaviors are observed in the predictions for $d\sigma/dY |_{y=0}$ show in Table \ref{tab1}.

\begin{table}[t]
\begin{center}
\begin{tabular} {||c|c|c|c|c|c|c||}
\hline
\hline
& {\bf HEAVY ION}   & {\bf MESON} & No shadowing & DS & EKS & EPS  \\
\hline
\hline
{\bf RHIC} & AuAu & $J/\Psi$ & 363 $\mu$b (113 $\mu$b) & 358 $\mu$b (108 $\mu$b)& 383 $\mu$b (83 $\mu$b)& 304 $\mu$b (53 $\mu$b) \\
\hline
\hline
 {\bf LHC}&  PbPb &  $J/\Psi$&  74 mb &  61 mb & 39 mb & 13 mb \\
 \hline
  &  & $\Upsilon$ &  163 $\mu$b &  148 $\mu$b & 120 $\mu$b & 22 $\mu$b \\
\hline
\hline
\end{tabular}
\end{center}
\caption{ The differential cross section at midrapidity ($d\sigma/dY |_{y=0}$) for nuclear vector meson photoproduction in UPC's at RHIC and LHC energies. At RHIC we also present in brackets the predictions  considering the nuclear breakup probability.}
\label{tab1}
\end{table}

Recently, the PHENIX collaboration has 
released the first (preliminary) data on the differential cross section of the coherent $J/\Psi$ production at mid-rapidity in gold - gold ultraperipheral collisions at $\sqrt{s} = 200$ GeV accompanied by $Au$ breakup \cite{phenix}. The value measured was: $d\sigma/dY|_{Y = 0} = 48. \pm 14. \mbox{(stat)} \pm 16. \mbox{(syst)} \, \mu b$. In order to compare with this  PHENIX data  we scale down our predictions by the nuclear breakup probability $P_{Xn} \approx 0.64$, which takes into account  the reduction of the yield expected when requiring coincident forward neutron emission. It corresponds to $J/\Psi$ production in coincidence with Coulomb breakup of at least one of the nuclei (For a detailed discussion see Refs. \cite{nystrand,baltz}). 
When compared with the preliminary PHENIX data we have that it is reasonably described by the EPS prediction (See Table \ref{tab1}). Although it is a very interesting result, since the typical values of $x$ probed at mid-rapidity are similar to those at forward rapidities in central d+Au collisions, which have motivated the inclusion of a stronger shadowing in the EPS parameterization, the current uncertainties and low statistics preclude yet any detailed conclusion. As shown in Ref. \cite{david} several models, based on very distinct assumptions, also describe this data. However, differently from the models proposed in Refs. \cite{klein_prc,strikman_plb,per4,ivanov_kop}, the EPS result predicts an enhancement of the rapidity distribution at large $|Y|$, which can be considered a signature of this model.


\begin{figure}[t]
\includegraphics[scale=0.35]{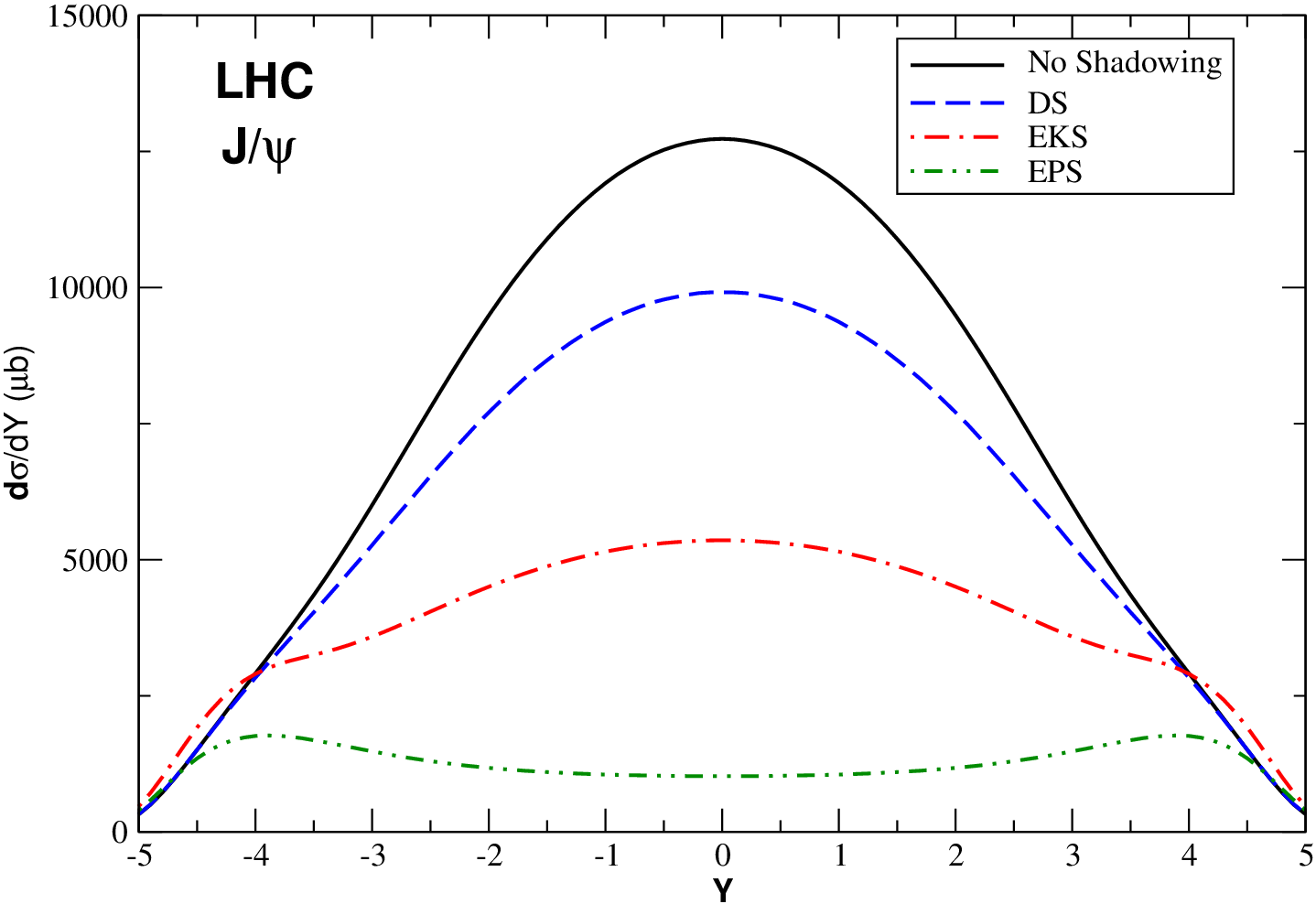}
\hspace{0.35cm}
\includegraphics[scale=0.35]{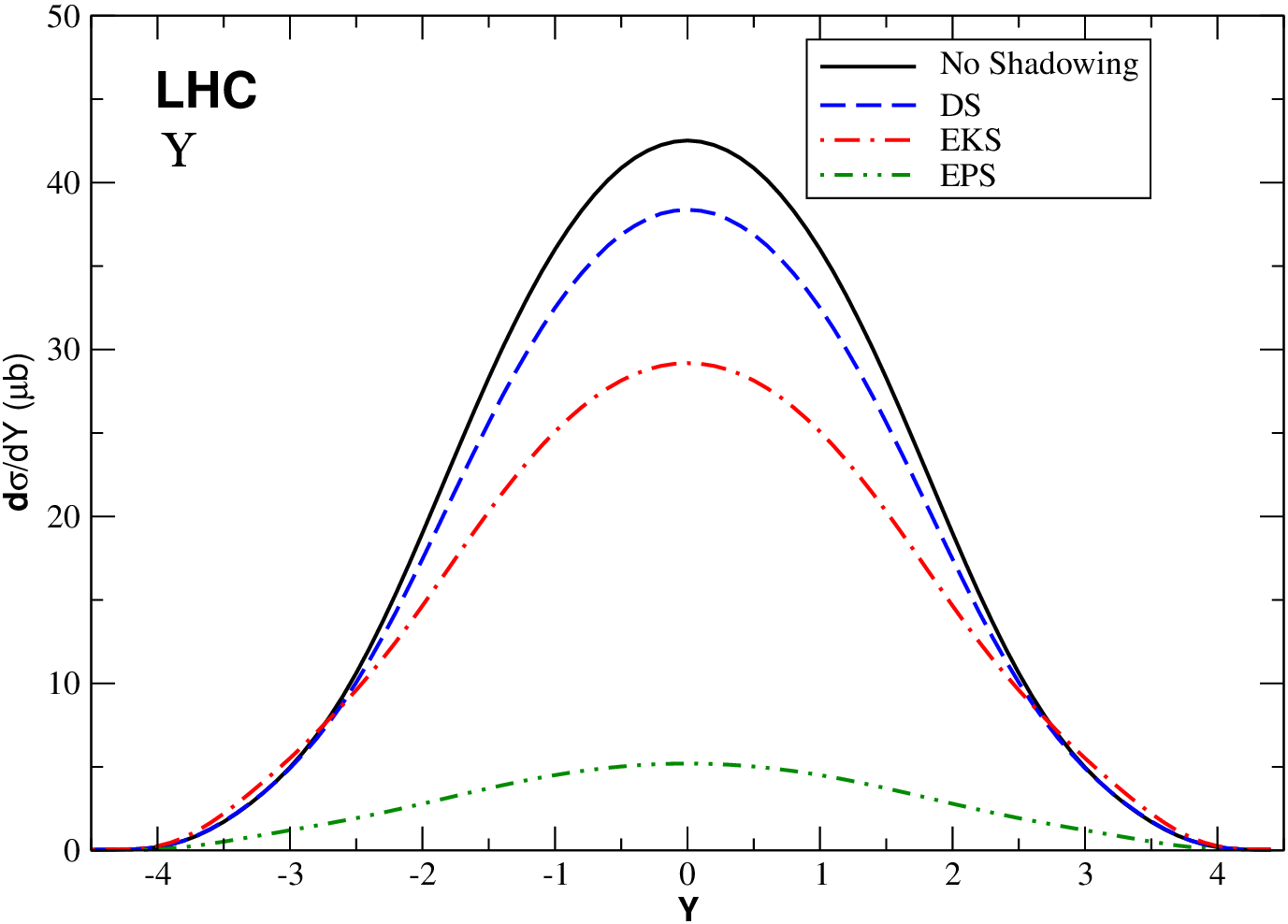}
\caption{(Color online) The rapidity distribution for nuclear vector meson  photoproduction on UPC's in $A A$  reactions at LHC  energy ($\sqrt{S_{NN}}=5.5\,\mathrm{TeV}$).}
\label{fig:lhc}
\end{figure}

In Fig. \ref{fig:lhc}  we present  our predictions for $J/\Psi$ (left panel) and $\Upsilon$ (right panel) production in coherent $AA$ collisions, considering $A = Pb$ and $\sqrt{s_{NN}} = 5.5$ TeV.
The production at mid-rapidity at the LHC probes $x$-values of order  $(6-20)\times 10^{-4}$, where the nuclear parameterizations differ by a factor 4 (See Fig. \ref{fig:rg}). The no shadowing prediction is presented for comparison. At LHC energies the rapidity distribution is strongly dependent of the magnitude of the shadowing effects in the nuclear gluon distribution. While the DS prediction implies a small reduction at mid-rapidity  in comparison to the no shadowing one, the rapidity distribution is suppressed by a factor 10 if calculated using the EPS gluon. Moreover, the difference between the predictions of the  three parameterizations is very large, which implies that the rapidity distribution for exclusive quarkonium photoproduction  in UPC's at LHC could be used to constrain the magnitude of the shadowing effect.

Finally, in Table \ref{tab1} we present our predictions for the integrated cross section at RHIC and LHC energies considering as input the distinct nuclear gluon distributions. In the RHIC case we present our results considering the suppression associated to the probability of nuclear breakup. In this case the no shadowing and DS predictions for the total cross section are very similar, as expected from Fig. \ref{fig:rhic}, while the EKS one predicts an enhancement of 10 \% and the EPS one a suppression of   20 \%. On the other hand, at LHC we have that the total cross section is strongly suppressed by the shadowing effects. For example, in $J/\Psi$ ($\Upsilon$) production, the EPS prediction is a factor $\approx 5\,(8)$ smaller than the no shadowing one.

In what follows we compare our results for the integrated cross sections with other theoretical predictions currently available.
The photoproduction of $J/\Psi$ in UPC's  was estimated in Refs. \cite{klein_prc,strikman_plb,per4,vicmag_prd2008,ivanov_kop}. In general our predictions at RHIC energies are larger than the results presented in those references when the nuclear breakup is included, with our EPS prediction being almost identical that shown in Ref. \cite{per4}.
This result is directly associated to the inclusion of the antishadowing effects in our predictions and/or due to the different treatment of the photon-nucleus  interaction.  At LHC energies, our predictions are similar those presented in Refs. \cite{klein_prc,strikman_plb,per4,ivanov_kop,vicmag_prd2008,bert4}. In particular, our no shadowing prediction is almost identical to  the impulse approximation prediction presented in Table 2 from Ref. \cite{bert4}. Moreover, 
our EPS prediction is almost identical  to the result obtained in Ref. \cite{strikman_plb} using a leading twist model of nuclear shadowing, while our EKS prediction is similar to the results presented in Refs. \cite{klein_prc,per4,vicmag_prd2008}. On the other hand, the photoproduction of $\Upsilon$ in UPC's  at LHC energies was estimated in Refs. \cite{strikman_jhep,klein_prl,vicmag_prd2008}. Our no shadowing prediction is almost identical to that presented in Ref. \cite{klein_prl}, where the nuclear effects are disregarded. On the other hand, our EPS prediction is smaller by a factor $\gtrsim 3$ than the results presented in Refs. \cite{strikman_jhep,klein_prl,vicmag_prd2008}, with the EKS one being larger by a factor $\gtrsim 1.25$ than the predictions from Refs. \cite{strikman_jhep,vicmag_prd2008}.
These results indicate that the study of $\Upsilon$ production can be useful to discriminate between the different models.

\begin{table}[t]
\begin{center}
\begin{tabular} {||c|c|c|c|c|c|c||}
\hline
\hline
& {\bf HEAVY ION}   & {\bf MESON} & No shadowing & DS & EKS & EPS  \\
\hline
\hline
{\bf RHIC} & AuAu & $J/\Psi$ & 363 $\mu$b & 358 $\mu$b & 383 $\mu$b & 304 $\mu$b \\
\hline
\hline
 {\bf LHC}&  PbPb &  $J/\Psi$&  74 mb &  61 mb & 39 mb & 13 mb \\
 \hline
  &  & $\Upsilon$ &  163 $\mu$b &  148 $\mu$b & 120 $\mu$b & 22 $\mu$b \\
\hline
\hline
\end{tabular}
\end{center}
\caption{ The integrated cross section for nuclear vector meson photoproduction at UPC's at RHIC and LHC energies.}
\label{tab2}
\end{table}

As a summary, in this paper we have calculated the rapidity distribution and
integrated cross sections of  exclusive photonuclear production of
vector mesons in ultraperipheral heavy ion collisions within the QCD collinear approach in the leading logarithmic approximation.
The main characteristic of  this approach is the quadratic dependence on the gluon distribution,
which makes it an excellent probe of the behavior of this
distribution.
We demonstrate that this model  reasonably describes  the current data on vector meson
photoproduction in scattering on protons. We extend this model for $\gamma A$ interactions and estimate  the cross sections for the $A+A
\rightarrow A+A+V$ ($V =  J/\Psi, \,\Upsilon$) process for RHIC and LHC energies considering as input different parameterizations for the nuclear gluon distribution.
We demonstrate that the rapidity distribution and the integrated cross section are strongly dependent on the model considered for the nuclear effects. It implies that these quantities
are powerful observables to discriminate among the different parton distributions in the nuclear medium. In particular,  the predicted shadowing by the EPS parameterization is considerably larger than in the previous nuclear PDF's. This strong shadowing effect in the quarkonium photoproduction  could be tested at the forthcoming LHC   through the detection of two rapidity gaps with the lepton pair  from the meson decay identified in the ATLAS, ALICE and CMS detectors. Our main conclusion it that the exclusive quarkonium photoproduction in ultraperipheral heavy ion collisions is a useful tool to help in obtaining the correct gluon distribution in the nuclear medium.

\section*{Acknowledgments}
 VPG have benefited from fruitful discussions with C. A. Bertulani and M. V. T. Machado.  This work was partially financed by the Brazilian funding agencies CNPq and FAPERGS.

\end{document}